# A variational approach for the deformation of a saturated porous solid. A second-gradient theory extending Terzaghi's effective stress principle

F. dell'Isola, M. Guarascio, K. Hutter

**Summary** The principle of virtual power is used to derive the equilibrium field equations of a porous solid saturated with a fluid, including second density-gradient effects; the intention is the elucidation and extension of the effective stress principle of Terzaghi and Fillunger. In the context of a first density-gradient theory for a saturated solid we interpret the porewater pressure as a Lagrange multiplier in the expression for the deformation energy, assuring that the saturation constraint is verified. We prove that this saturation pressure is distributed among the constituents according to their respective volume fraction (Delesse law) only if they are both true density-preserving. We generalize the Delesse law to the case of compressible constituents. If a material-dependent effective stress contribution is to arise, it is, in general, nonvanishing simultaneously in both the solid and fluid constituents. Moreover, saturation pressure, effective stresses and compressibility constitutive equations determine the exchange volume forces. In a theoretical formulation without non-isotropic strain measures, second density-gradient effects must be incorporated, not only to accommodate for the equilibrium-solid-shear stress and the interaction among neighboring solid-matrix pores, but also to describe internal capillarity effects. The earlier are accounted for by a dependence of the thermodynamic energy upon the density-gradient of the solid, while the latter derives from a mixed density-gradient dependence. Examples illustrate the necessity of these higher gradient effects for properly posed boundary value problems describing the mechanical behaviour of the disturbed rock zone surrounding salt caverns. In particular, we show that solid second-gradient strains allow for the definition of the concept of static permeability, which is distinct from the dynamic Darcy permeability.

## 1 Introduction

A saturated porous material is a mixture of a solid matrix with pores which are completely filled with a fluid, called the porewater fluid. The state of stress in a material point of this mixture consists of two contributions; one contribution is the constraint pressure $p$ due to saturation, called in modern porous media theories the *saturation pressure,* in the soil science literature, however, better known as the *porewater pressure.* This pressure is distributed between the two constituents: in older theories via a postulate (Delesse law, s. e.g. [1]), according

F. dell'Isola
Dipartimento di Ingegneria, Strutturale e Geotecnica, Università di Roma La Sapienza, Via Eudossiana 8, I-00184 Roma, Italia

M. Guarascio
Dipartimento di Ingegneria Chimica, dei Materiali, delle Materie Prime e Metallurgia, Università di Roma, La Sapienza
Via Eudossiana 8, I-00184 Roma, Italia

K. Hutter
Department of Mechanics, Darmstadt University of Technology,
D-64289 Darmstadt, Germany

to volume fraction occupied by the respective constituents, $p_a = pv_a$, $v_a$ is the volume fraction of constituent $a$; in newer theories by derivation. The other contribution to the stresses is constitutive, and thus responsible for the strength of the material. It is customary in the solid mechanics literature to refer to these second types of stresses as the *effective stresses* (note plural) in the solid and fluid, respectively. In anticipation that the fluid is true density-preserving, the solid carries all constitutive properties, and the term *effective stress* is only used in the singular form referring to the solid matrix.

This understanding was that of Terzaghi and Fillunger, and the latter was the first soil engineer who clearly stated that a constitutive equation must be formulated only for the excess over the weighted porewater pressure and not for the total stress, s. [1] p. 81–83. However, also Terzaghi was of the opinion that the excess over the porewater stress "has its seat exclusively in the solid phase of the soil" (according to [1] p. 83, quoting Skempton 1960). Moreover, both Fillunger and Terzaghi postulate that solid volume fraction kinematically describes the strain of the solid matrix.

According to the above statements three questions are raised and will be answered in this paper: (1) Is the division of the porewater pressure between the solid and fluid constituents truly according to "pressure equilibrium", $p_a = pv_a$, or can it be different? (2) Is the concept of *effective stress* one that must be restricted to the solid matrix as expected by Terzaghi or does it apply for the solid *and the fluid*? (3) Is the solid volume fraction sufficient to describe the solid volume pore strain?

To answer these questions we will use the principle of virtual power applied to a binary mixture of a solid saturated with a fluid. In order to simplify the required mathematical formalism we will limit ourselves to consider equilibrium configurations of the considered mixture. The flexibility in the underlying postulates will be assumed to be sufficiently broad to cover the conjectured properties. We will assume, therefore, that the two constituents may be either compressible or true density-preserving, i.e. both situations can be imposed by specialization. Second, we shall assume the thermodynamic energy to depend both on the partial densities of the constituents as well as on their gradients. Third, by postulating the thermodynamic energy to depend on the density gradients an incompleteness of the solid matrix strain description will be removed. The latter dependence will give rise to double gradients of the densities in the constitutive relations for the pressure, which is why dependence on the density gradients in the energy is called a *second-gradient theory*. Gradients higher than the first are e.g. needed to transmit nonspherical equilibrium stresses. In the application developed in the last Section we show that second-gradient strains of the solid matrix allow the definition of a static permeability, i.e. of a physical quantity measuring how under equilibrium conditions (in the absence of fluid motion) a porous solid matrix offers resistance to the variation of the saturating fluid volume fraction. Other dependencies could be incorporated but the intention is to keep the theory as simple as possible.

The first two questions raised above and modifying the Terzaghi-Fillunger conjectures can be answered with a first-gradient theory. It will be shown that, within the context of this variational formulation, the porewater pressure is, in general, not distributed between the constituents according to pressure equilibrium, as postulated by the Delesse law, and effective stresses arise for both the solid and the fluid constituents. Moreover, in order that the equilibrium equations derived from the variational principle be consistent mixture balances, the constitutive relations for the thermodynamic energy, the effective stresses, the constituent compressibilities and exchange volume force between constituents must be related by imposing objectivity of the power expended by internal contact actions [2, 3]. The resultant relations (s. the following Eqs. (20)–(22)) can be used in the framework of the present variational formulation to prove the following statements:

- When both constituents are true density-preserving, the porewater pressure is distributed among the constituents according to the "pressure equilibrium" postulate (Delesse law). On the other hand, when the constituents are compressible, the Delesse law is no longer valid, and the porewater pressure is distributed according to a prescribed formula.
- Fillunger's [25] statement that compressibility leads to a constitutive effective stress is corroborated.
- The interaction force has a contribution proportional to the gradient of the solid volume fraction. The coefficient is, in general, not equal to the porewater pressure, but reduces to the latter where the effective pressures vanish.

- Effective stresses, in general, arise simultaneously in both constituents, and are related to the mixture thermodynamic pressure. This statement is also in agreement with other formulations of granular solid-fluid mixtures such as [4][1] or [6].
- The second gradient effects must be incorporated in a theory ignoring a dependence of the thermodynamic quantities on nonhydrostatic strain measures or rate-independent plastic effects, in order to allow for equilibrium shear stresses. These effects are also needed to describe the effects on stress in a material point related to the solid-matrix-pore deformation in the neighbouring material points.

These results seem to be new. One advantage of the variational formulation over direct approaches to higher gradient theories is the fact that the minimization principle automatically generates the natural boundary conditions. In other words, not only the dynamic equations within the body relating the introduced fields are generated, but also the conditions which these fields must fulfill at the boundaries of the body. In first-gradient theories these boundary conditions lend themselves easily through arguments of physics, for higher-gradient theories they must fall out from the formulation, because this physical intuition is generally not available. We adapt the arguments developed by [2, 3, 7–9], in order to physically interpret for the considered system the deduced boundary conditions. Furthermore, we demonstrate their appropriateness by solving a simple one-dimensional boundary value problem. It catches, at least qualitatively, some physical properties, [10], of deformable salt rocks fully saturated by a fluid, which cannot be described without the introduction of second density gradients.

## 2
## Conceptual prerequisites

Consider a binary mixture of a solid matrix with connected pores which are filled with a liquid. This arrangement can be thought of as being a saturated soil or rock. We have in mind as an application, however, a dome of salt which is saturated with a brine of given concentration. Solidification of salt from the brine to the solid matrix is excluded. Let the two components be referred to as the solid and the porewater fluid and be indicated by the suffices $s$ and $f$. Let, moreover, $\rho_s$, $\rho_f$ and $\mathbf{v}_s, \mathbf{v}_f$ be the solid and fluid densities and velocities, respectively, in the mixture. The mixture density and the barycentric velocity are then given by

$$\rho = \rho_s + \rho_f = \hat{\rho}_s v_s + \hat{\rho}_f v_f = \hat{\rho}_s v_s + \hat{\rho}_f (1 - v_s) \ , \tag{1}$$

$$\mathbf{v} = \frac{\rho_s}{\rho}\mathbf{v}_s + \frac{\rho_f}{\rho}\mathbf{v}_f =: \xi_s \mathbf{v}_s + \xi_f \mathbf{v}_f \ , \tag{2}$$

in which $v_s$ is the solid volume fraction, $v_f$ the porosity; the saturation condition has been used stating that the pore-fluid fills the entire pore space, and $\hat{\rho}_s$, $\hat{\rho}_f$ are the true densities of the solid and the fluid.

We conceive this mixture to be nonreactive so that the balances of mass for the constituents reduce to

$$\frac{\partial \rho_a}{\partial t} + \nabla \cdot (\rho_a \mathbf{v}_a) = 0, \quad a = s, f \ . \tag{3}$$

In the ensuing analysis we shall restrict ourselves to purely mechanical processes; temperature will play no role, and so the constituent momentum equations are the only additional balance laws to be added to (3). Instead of a direct application of these laws, we shall use to derive them the principle of virtual power, applied to the appropriate energy functional. Let $\psi$ be this functional, and assume it to depend on the fields $\rho_a$ and $\nabla \rho_a$, $a = s, f$,

$$\psi = \psi(\rho_a, \nabla \rho_a) \ . \tag{4}$$

Other dependences could also be introduced, e.g. a temperature dependence and dependences on tensorial strain measures; however, our interest is in the derivation of the most simple Terzaghi-type effective stress theory including second gradient effects. Imposing objectivity, Eq. (4) takes the form

$$\psi = \psi(\rho_s, \rho_f, |\nabla \rho_s|^2, |\nabla \rho_f|^2, |\nabla \rho_s \cdot \nabla \rho_f|) \ . \tag{5}$$

---

[1] In the formulation of [5], the fluid stresses are absent in the reduced formulation, corresponding to this one due to a slip in calculations.

This form of the energy functional supposes that both the solid matrix as well as the fluid exhibit second-gradient effects, and that the two also give rise to an interaction energy of the two effects through the last variable in (5).

The principle of virtual power states that the variation of the total energy in the body related to its motion equals the power of the external forces, i.e.

$$\frac{d}{dt}\left(\int_{\mathfrak{B}} \psi dV\right) = \int_{\mathfrak{B}} (\mathbf{b}_s \cdot \mathbf{v}_s + \mathbf{b}_f \cdot \mathbf{v}_f)dV + \int_{\partial \mathfrak{B}} (\mathfrak{C}_s(\mathbf{v_s}, \nabla \mathbf{v}_s) + \mathfrak{C}_f(\mathbf{v}_f, \nabla \mathbf{v}_f))dS \ . \tag{6}$$

Here, $\mathbf{b}_a$ are the specific body forces which perform work on their constituent motions, indicated by the velocity fields $\mathbf{v}_s$ and $\mathbf{v}_f$, respectively, while integrals of $\mathfrak{C}_a$ are scalar-valued bilinear functionals. They represent the power expended on the velocity fields and their gradients by specific contact actions; this is so because, as in every second-gradient theory (s. e.g. [2, 3] or [9]), contact forces exert power also on gradients of velocity fields. The variation of the total energy on the left-hand side of (6) is

(i) referred to a barycentric motion in the following Sec. 3, and to constituent motions when considering the second gradient terms added in Sec. 5;
and
(ii) calculated by independently varying the independent variables of the energy function subject to the kinematic constraint that the body is saturated.

The latter condition is now incorporated by defining the energy according to

$$\psi = \varepsilon + p(\nu_s + \nu_f - 1) = \varepsilon(\rho_a, \nabla \rho_a) + p\left(\frac{\rho_s}{\hat{\rho}_s} + \frac{\rho_f}{\hat{\rho}_f} - 1\right) \ , \tag{7}$$

in which $\hat{\rho}_a$ are the true peculiar densities, assumed to be constants for density-preserving constituents; note that $\rho_a = \nu_a \hat{\rho}_a$, and $p$ is a Lagrange multiplier which will be determined by imposing the saturation constraint; the quantity $\varepsilon$ is the thermodynamic free energy. In the case that entropy is the independent thermodynamic field variable, $\varepsilon$ is the internal energy. If this independent field variable is the temperature, then $\varepsilon$ is the Helmholtz free energy. Because we do not specify this, we are either concerned with isentropic or isothermal processes. With the interpretation (7), the variation in (6) can be performed for unconstrained $\rho_a$.

Finally, let us state that we do not require *ab initio* that the components be incompressible. In fact, it will be supposed that the true densities $\hat{\rho}_a$ may be affected by the composition of the grains. At high porosity, $\hat{\rho}_a$ will essentially be constant, at low porosity near the closest packing, $\hat{\rho}_a$ will itself increase. This dependence was clearly noted by Fillunger, and can be accounted for by postulating a constitutive relation $\hat{\rho}_a = \hat{\rho}_a(\rho_a)$. For the final application, we will assume that $\hat{\rho}_f = $ const. and $\hat{\rho}_s = \hat{\rho}_s(\rho_s)$.

## 3
## The classical model

Consider, as a preliminary approach, a theory without second gradient effects. Then

$$\psi = \varepsilon(\rho_b) + p\left(\frac{\rho_s}{\hat{\rho}_s} + \frac{\rho_f}{\hat{\rho}_f} - 1\right) \ , \tag{8}$$

thus

$$\frac{d}{dt}\left(\int_{\mathfrak{B}} \psi dV\right) = \int_{\mathfrak{B}} \mathfrak{L}\psi dV \ , \tag{9}$$

where

$$\mathfrak{L}\psi := \frac{\partial \psi}{\partial t} + \nabla \cdot (\psi \mathbf{v}) \ .$$

With

$$\begin{aligned}\frac{\partial \psi}{\partial t} &= \sum_a \frac{\partial \psi}{\partial \rho_a}\frac{\partial \rho_a}{\partial t} \stackrel{(3)}{=} -\sum_a \frac{\partial \psi}{\partial \rho_a}\nabla \cdot (\rho_a \mathbf{v}_a), \\ \nabla \cdot (\psi \mathbf{v}) &= \sum_a (\psi \xi_a \nabla \cdot \mathbf{v}_a + \psi \nabla \xi_a \cdot \mathbf{v}_a + \nabla \psi \cdot \xi_a \mathbf{v}_a) \ ,\end{aligned} \tag{10}$$

it is readily shown that

$$\int_\mathfrak{B} \mathfrak{L}\psi dV = \sum_a \int_\mathfrak{B} [-p_a \nabla \cdot \mathbf{v}_a + (\psi \nabla \xi_a - \frac{\partial \psi}{\partial \rho_a}\nabla \rho_a) \cdot \mathbf{v}_a + \nabla \psi \cdot \xi_a \mathbf{v}_a] dV \ , \qquad (11)$$

where

$$p_a := -\psi \xi_a + \rho_a \frac{\partial \psi}{\partial \rho_a} \ .$$

The first term in the integrand of (11) can be transformed according to

$$\begin{aligned}\int_\mathfrak{B} -p_a \nabla \cdot \mathbf{v}_a dV &= -\int_\mathfrak{B} \nabla \cdot (p_a \mathbf{v}_a) dV + \int_\mathfrak{B} \nabla p_a \cdot \mathbf{v}_a dV \\ &= -\int_{\partial \mathfrak{B}} p_a \mathbf{v}_a \cdot \mathbf{n} dS + \int_\mathfrak{B} \nabla p_a \cdot \mathbf{v}_a dV \ .\end{aligned} \qquad (12)$$

With (11) and (12), the principle of virtual work (6), takes the form

$$\begin{aligned}&\sum_a \int_{\partial \mathfrak{B}} -p_a \mathbf{n} \cdot \mathbf{v}_a dS + \sum_a \int_\mathfrak{B} \left(\nabla\left(\rho_a \frac{\partial \psi}{\partial \rho_a}\right) - \frac{\partial \psi}{\partial \rho_a}\nabla \rho_a\right) \cdot \mathbf{v}_a dV \\ &= \sum_a \int_\mathfrak{B} \mathbf{b}_a \cdot \mathbf{v}_a dV + \sum_a \int_{\partial \mathfrak{B}} \mathfrak{C}_a(\mathbf{v}_a, \mathbf{0}) dS \ .\end{aligned} \qquad (13)$$

This must hold for all fields $\mathbf{v}_a$ defined over the body $\mathfrak{B}$ and on its boundary $\partial \mathfrak{B}$, thus leading to

$$-\rho_a \nabla \left(\frac{\partial \psi}{\partial \rho_a}\right) + \mathbf{b}_a = 0, \quad a = s, f \text{ in } \mathfrak{B} \ , \qquad (14)$$

$$-p_a \mathbf{n} + \mathfrak{C}_a(\mathbf{v_a}, \mathbf{0}) = 0, \quad a = s, f \text{ on } \partial \mathfrak{B} \ . \qquad (15)$$

In order to identify in the last equation the addends representing partial stresses and exchange bulk force between the constituents, we impose the same Galilean invariance argument as done in [2, 3] and [7]: it follows that

$$\begin{aligned}-\nabla p_a + \mathbf{m}_a + \mathbf{b}_a &= 0, \quad a = s, f \text{ in } \mathfrak{B}, \\ -p_a \mathbf{n} + \mathbf{t}_a &= 0, \quad a = s, f \text{ in } \partial \mathfrak{B} \ ,\end{aligned} \qquad (16)$$

in which the exchange bulk forces $\mathbf{m}_a$ are given by

$$\mathbf{m}_s = -\mathbf{m}_f = -(1-\xi_s)\nabla \rho_s \left(\frac{\psi}{\rho} - \frac{\partial \psi}{\partial \rho_s}\right) + (1-\xi_f)\nabla \rho_f \left(\frac{\psi}{\rho} - \frac{\partial \psi}{\partial \rho_f}\right) \ , \qquad (17)$$

and $\mathbf{t}_a$ is the linear part of $\mathfrak{C}_a(\mathbf{v}_a, \mathbf{0})$ on $\mathbf{v}_a$ and represents a stress vector.

We remark explicitly that when the saturation constraint is not considered, and therefore $p$ in Eq. (8) vanish, Eq. (17) coincides with Eq. (4.6) in [6].

With $\rho_a = v_a \hat{\rho}_a$, $\hat{\rho}_a = \hat{\rho}_a(\rho_a)$, and $\psi$ as given by (8), since $\nabla v_f = -\nabla v_s$ straightforward arithmetics shows that

$$\hat{\rho}_a \nabla v_a = \nabla \rho_a \left(1 - v_a \frac{d\hat{\rho}_a}{d\rho_a}\right), \quad \rho_a \frac{\partial \psi}{\partial \rho_a} = v_a \left[\hat{\rho}_a \frac{\partial \varepsilon}{\partial \rho_a} + p\left(1 - v_a \frac{d\hat{\rho}_a}{d\rho_a}\right)\right] \ , \qquad (18)$$

$$p_a = -\rho_a \frac{\psi}{\rho} + \rho_a \frac{\partial \hat{\psi}}{\partial \rho_a} = \mathscr{P}_a + p v_a \left(1 - v_a \frac{d\hat{\rho}_a}{d\rho_a}\right) \ , \qquad (19)$$

$$\mathbf{m}_a = M \nabla v_a, \quad M := p + \sum_b \frac{\mathscr{P}_b}{v_b}(1 - \xi_b)\left(1 - v_b \frac{\mathrm{d}\hat{\rho}_b}{\mathrm{d}\rho_b}\right)^{-1} . \tag{20}$$

We cannot resist to point out here the close similarity between Eq. (19) and the well-known formula

$$\sigma_m^{\mathrm{eff}} = \sigma_m + \left(1 - \frac{k_0}{k_s}\right) P , \tag{21}$$

of soil mechanics (s. [11–13]), with the correspondence

$$\sigma_m^{\mathrm{eff}} \leftrightarrow -\mathscr{P}_a \qquad \sigma_m \leftrightarrow -p_a \qquad P \leftrightarrow p v_a ,$$

where $\sigma_m^{\mathrm{eff}}$ is the mean effective stress, $\sigma_m$ the mean applied stress and $P$ the pore pressure. The factor of the second term on the right-hand side of (21) is called the Biot coefficient, and $k_0, k_s$ are the moduli of compressibility of the drained soil and the rock material. It is tempting to identify $k_0/k_s$ with $v_a \mathrm{d}\hat{\rho}_a/\mathrm{d}\rho_a$, but this is no more than a suggestion.

The expression (19) holds modulo $p(\sum_a v_a - 1)$, which is zero, and the quantity

$$\mathscr{P}_a := \rho_a \left(\frac{\partial \varepsilon}{\partial \rho_a} - \frac{\varepsilon}{\rho}\right) \tag{22}$$

may be identified with the thermodynamic pressure of the constituent $a$. Therefore, the equilibrium equation $(16)_1$ takes the form

$$\begin{array}{ccc}(1) & (2) & (3)\\ -\nabla p_a & +M\nabla v_a & +\mathbf{b}_a\end{array} = \mathbf{0} . \tag{23}$$

Notice that Eq. (23) expresses a balance between the pressure gradient (1), the exchange force (2) and the external body force (3), which necessarily must satisfy the basic postulates of the mixture balance laws. We also mention that, for $\mathrm{d}\hat{\rho}_b/\mathrm{d}\rho_b = 0$, the above expression for $\mathbf{m}_a$ coincides with that found with different methods in [4]. The quantity $\mathscr{P}_b/v_b$ was denoted by $\beta_b$ in [5].

Relation (20) serves as a restriction among constitutive equations: for instance, if $\hat{\rho}_f$ and $\hat{\rho}_s$ are both constant, i.e. if the two constituents are true density-preserving, then (20) requires that, once the constitutive quantity $\varepsilon$ and the pressure $p$ are assigned, so is the exchange force $M$. In particular, if $\varepsilon = 0$ and $\mathrm{d}\hat{\rho}_b/\mathrm{d}\rho_b = 0$, Eq. (23) reduces to

$$-\nabla(v_a p) + p\nabla v_a + \mathbf{b}_a = \mathbf{0} , \tag{24}$$

or

$$-v_a \nabla p + \mathbf{b}_a = \mathbf{0} . \tag{25}$$

The first form shows that the partial pressures $p_a = p v_a$ are obtained from the saturation (porewater) pressure by multiplying it with the constituent volume fraction, a property usually referred to as "pressure equilibrium". On the other hand, the interaction force is simply given by $-p\nabla v_a$. Such choices are the basis of a large number of porous solid theories (s. e.g. [14–16]), but this choice is obviously very restrictive.

The above results show that

1. Fillunger [25] was essentially correct when conjecturing that an extension of the classical theory exhibited by (24) can be obtained via compressibility assumptions of the constituents solid and fluid, (s. [1], p. 82, footnote 8): indeed Eq. (22) define the Filllunger–Terzaghi effective partial pressures, and show that they are related to the compressibility constitutive relations $\hat{\rho}_b(\rho_b)$.

2. Both Fillunger and Terzaghi (s. again [1]) did not investigate the implications of the balance of energy and of the second principle of the thermodynamics in the theory which they were developing; in particular, they assumed that (using the notation of the present paper)

$$\mathscr{P}_s \neq 0, \qquad \mathscr{P}_f = 0, \qquad M = p , \tag{26}$$

which clearly contradicts via (20) the principle of virtual powers on which we have based our treatment.

## 4
## Limits of the classical model

By adding over the constituents their force balances (23) an expression for the porewater pressure can be found. Indeed, by assuming that the body forces are conservative and possess a potential

$$\sum_a \mathbf{b}_a = -\nabla \phi \ , \tag{27}$$

it is straightforward to show that

$$\nabla \left\{ \sum_a \left[ \left(1 - v_a^2 \frac{\mathrm{d}\hat{\rho}_a}{\mathrm{d}\rho_a}\right) p + \mathscr{P}_a \right] + \phi \right\} = 0 \ , \tag{28}$$

or, after integration,

$$p \left(1 - \sum_a v_a^2 \frac{\mathrm{d}\hat{\rho}_a}{\mathrm{d}\rho_a}\right) + \sum_a \mathscr{P}_a + \phi = k \ , \tag{29}$$

where $k$ may be a function of time, which we take to be constant. This result shows that for homogeneous fluid-saturated solid matrices and for a vanishing potential field $\phi$ (i.e. $\mathbf{b}_a = 0$ for $a = s, f$), the pressure field $p$ is constant if and only if one of the fields $\rho_a$ or $v_a$ is constant. On the other hand, it is easy to verify that, under the same hypothesis on $\phi$, both Eq. (23) are solved by constant fields $\rho_a$, as can be checked immediately with the aid of (15). These results hold, no matter what the geometry of the solid matrix may be, and irrespective of whether the body be in equilibrium or in a dynamic Stokesian motion.

On the other hand, it is easy to conceive physical situations in which a solid matrix saturated by a fluid shows a variable solid pore size distribution even in an equilibrium configuration, and, consequently, a spatially variable solid-volume fraction. An instance of a physical system in which this occurs is given by the micro-cracked and permeable salt rock in the disturbed zone surrounding a fluid filled cavity, [17]. It seems, therefore, useful to formulate a model in which the micro-mechanical interactions among neighbouring pores are described in such a way that the existence of a spatially varying equilibrium solid volume fraction field is permissible in the absence of body forces. This will be done in the next Section, by introducing a second gradient theory.

## 5
## A second-gradient theory

We choose now the thermodynamic energy in the form

$$\varepsilon = \varepsilon(\rho_a, f_{ab}) = \varepsilon(\rho_a, f_{ab}), \quad f_{ab} := \nabla \rho_a \cdot \nabla \rho_b \ , \tag{30}$$

and thus need to complement the variation of $\int_{\mathfrak{B}} \psi \mathrm{d}V$ performed in Secs. 2 and 3 only by the contributions due to the additional term involving the variables $f_{ab}$. Indicating this contribution by the index $()_{add}$, we may write

$$\left(\frac{\mathrm{d}}{\mathrm{d}t}\int_{\mathfrak{B}} \varepsilon \mathrm{d}V\right)_{add} = \sum_{a,b} \int_{\mathfrak{B}} \left(\frac{\partial \varepsilon}{\partial f_{ab}} \nabla \rho_b \cdot \nabla \rho'_a\right) \mathrm{d}V \ , \tag{31}$$

in which $\nabla \rho'_a$ denote the time derivatives following the motion of the solid and fluid, respectively, viz.

$$()'_a = \frac{\partial ()_a}{\partial t} + \nabla ()_a \cdot \mathbf{v}_a \ . \tag{32}$$

Using (32) in (31) and employing the rule

$$-\nabla \rho_b \cdot \nabla \rho'_a = (\nabla \rho_a \otimes \nabla \rho_b + f_{ab}\mathbf{I}) : \nabla \mathbf{v}_b + \rho_b(\mathbf{I} \otimes \nabla \rho_a) \vdots \nabla\nabla \mathbf{v}_b , \qquad (33)$$

obtained by recalling the balances of mass for the constituents, where no summation over repeated indices is performed, we may derive the identity

$$\left(\frac{\mathrm{d}}{\mathrm{d}t}\int_{\mathfrak{B}}\varepsilon \mathrm{d}V\right)_{add} = -\sum_b \left( \int_{\mathfrak{B}} \left( \sum_a \frac{\partial \varepsilon}{\partial f_{ab}} (\nabla \rho_a \otimes \nabla \rho_b + f_{ab}\mathbf{I}) \right) : \nabla \mathbf{v}_b \mathrm{d}V \right.$$
$$\left. + \int_{\mathfrak{B}} \left( \sum_a \frac{\partial \varepsilon}{\partial f_{ab}} \rho_b (\mathbf{I} \otimes \nabla \rho_a) \right) \vdots \nabla\nabla \mathbf{v}_b \mathrm{d}V \right) . \qquad (34)$$

With the notations

$$\mathbf{A}_b \equiv \sum_a \frac{\partial \varepsilon}{\partial f_{ab}} (\nabla \rho_a \otimes \nabla \rho_b + f_{ab}\mathbf{I}) , \qquad (35)$$

$$\mathbf{C}_b \equiv \left( \sum_a \frac{\partial \varepsilon}{\partial f_{ab}} \rho_b (\mathbf{I} \otimes \nabla \rho_a) \right) , \qquad (36)$$

the right-hand side of (34) can be regarded as the sum of terms possessing the structure

$$\int_{\mathfrak{B}} (\mathbf{A}_b : \nabla \mathbf{v}_b + \mathbf{C}_b \vdots \nabla\nabla \mathbf{v}_b) \mathrm{d}V , \qquad (37)$$

which, through rearrangements of differentiations and the application of Gauss' theorem, may be written as (s. [18], p. 19)

$$\int_{\mathfrak{B}} \nabla \cdot (\mathbf{A}_b - \nabla \cdot \mathbf{C}_b) \cdot \mathbf{v}_b \mathrm{d}V - \int_{\partial\mathfrak{B}} (\mathbf{v}_b \cdot (\mathbf{A}_b - \nabla \cdot \mathbf{C}_b) + \nabla \mathbf{v}_b : \mathbf{C}_b) \cdot \mathbf{n} \mathrm{d}S . \qquad (38)$$

Adding the transformed Eq. (34) to the variation calculated in the previous Section we obtain

$$\sum_a \int_{\mathfrak{B}} \left[ \nabla \left( \rho_a \frac{\partial \psi}{\partial \rho_a} \right) - \frac{\partial \psi}{\partial \rho_a} \nabla \rho_a - \nabla \cdot (\mathbf{A}_a - \nabla \cdot \mathbf{C}_a) \right] \cdot \mathbf{v}_a \mathrm{d}V$$
$$+ \sum_a \int_{\partial\mathfrak{B}} [\mathbf{v}_a \cdot (p_a \mathbf{n} + \mathbf{A}_a - \nabla \cdot \mathbf{C}_a) + \nabla \mathbf{v}_a : \mathbf{C}_a] \mathrm{d}S$$
$$= \sum_a \int_{\mathfrak{B}} \mathbf{v}_a \cdot \mathbf{b}_a \mathrm{d}V + \sum_a \int_{\partial\mathfrak{B}} \mathfrak{C}_a(\mathbf{v}_a, \nabla \mathbf{v}_a) \mathrm{d}S . \qquad (39)$$

This identity proves that the expression on the right-hand side is a bilinear functional of the velocity field $\mathbf{v}_a$ and its gradient. Thus, we write

$$\mathfrak{C}_a(\mathbf{v}_a, \nabla \mathbf{v}_a) = \mathbf{t}_a \cdot \mathbf{v}_a + \mathfrak{d}_a : \nabla \mathbf{v}_a , \qquad (40)$$

and call $\mathbf{t}_a$ surface traction vector and $\mathfrak{d}_a$ surface double-force tensor of constituent $a$. Formula (40) is nothing else than Cauchy's theorem extended to the second-gradient mixture theory treated here.

Since (39) must be valid for all $\mathbf{v}_a$ and $\nabla \mathbf{v}_a$, we finally deduce the field equations

$$\nabla \left( \rho_a \frac{\partial \psi}{\partial \rho_a} \right) - \frac{\partial \psi}{\partial \rho_a} \nabla \rho_a - \nabla \cdot (\mathbf{A}_a - \nabla \cdot \mathbf{C}_a) + \mathbf{b}_a = 0, \quad \text{in } \mathfrak{B} , \qquad (41)$$

and the boundary conditions

$$[-p_a \mathbf{I} - (\mathbf{A}_a - \nabla \cdot \mathbf{C}_a)] \cdot \mathbf{n} = \mathbf{t}_a \atop \mathbf{C}_a \cdot \mathbf{n} = \mathfrak{d}_a \Bigg\} \quad \text{on } \partial \mathfrak{B} \ . \tag{42}$$

The surface traction vectors $\mathbf{t}_a$ and the surface contact double forces $\mathfrak{d}_a$ for the constituent stresses are to be externally prescribed at the boundary $\partial \mathfrak{B}$ (with a unit normal $\mathbf{n}$) of a mixture body. In a first-gradient theory, both $\mathbf{A}_a$ and $\mathbf{C}_a$ vanish, and Eqs. (42) reduce to (16)$_2$.

As an illustration, we choose $\partial \varepsilon / \partial f_{ab} = 0$, except for

$$\frac{\partial \varepsilon}{\partial f_{ss}} =: \frac{\lambda_s}{2} \ . \tag{43}$$

Then the field equations become

$$-\nabla p_s + \mathbf{m}_s + \mathbf{b}_s + \nabla \cdot \left[ \left( \lambda_s \rho_s \triangle \rho_s + \frac{\lambda_s}{2} |\nabla \rho_s|^2 \right) \mathbf{I} - \lambda_s \nabla \rho_s \otimes \nabla \rho_s \right] = 0,$$
$$-\nabla p_f + \mathbf{m}_f + \mathbf{b}_f = 0 \ , \tag{44}$$

and the boundary conditions take the form

$$\left( -p_s + \lambda_s \rho_s \triangle \rho_s + \frac{\lambda_s}{2} |\nabla \rho_s|^2 \right) \mathbf{n} - \lambda_s (\nabla \rho_s \otimes \nabla \rho_s) \cdot \mathbf{n} + \mathbf{t}_s = 0,$$
$$-p_f \mathbf{n} + \mathbf{t}_f = 0, \quad \lambda_s \rho_s \frac{\partial \rho_s}{\partial n} + \mathfrak{d} = 0 \ , \tag{45}$$

in which

$$\mathfrak{d}_f = \mathbf{0}, \quad \mathfrak{d}_s = \mathfrak{d} \mathbf{I} \ .$$

Following the results found in [19–21] for the constituent tractions, denoting $p_i$ the incumbent pressure exerted on the saturated solid matrix at its boundaries, we choose

$$\mathbf{t}_s = \alpha v_s^l p_i \mathbf{n}, \quad \mathbf{t}_f = \left( 1 - \alpha v_s^l \right) p_i \mathbf{n} \ , \tag{46}$$

where $\alpha$ and $l$ are constitutive parameters describing the dynamic behaviour of the solid-matrix fluid interface. Moreover, we assume that

$$\mathfrak{d} = \mathfrak{d}(p_i) \tag{47}$$

is a given function of the incumbent pressure, which in ensuing application will be taken as linear

$$\mathfrak{d} = \mathfrak{D} p_i \ .$$

Finally, it is easy to show that when Eqs. (44) are added we can generalize (28), and obtain

$$\nabla \left\{ \sum_a \left[ \left( 1 - v_a^2 \frac{d \hat{\rho}_a}{d \rho_a} \right) p + \mathscr{P}_a \right] + \phi \right\}$$
$$- \nabla \cdot \left[ \left( \lambda_s \rho_s \triangle \rho_s + \frac{\lambda_s}{2} |\nabla \rho_s|^2 \right) \mathbf{I} - \lambda_s \nabla \rho_s \otimes \nabla \rho_s \right] = 0 \ . \tag{48}$$

## 6
## Linearized constitutive equations

In this Section we consider the constitutive equations for the energy $\varepsilon$ and true densities $\hat{\rho}_a$ in the neighbourhood of an equilibrium configuration $(\rho_s^0, \rho_f^0)$ which is stress free. Having in mind the application to the study of equilibrium configurations of cracked salt rocks saturated with a fluid, the constitutive postulate (49) on $\varepsilon$ can be interpreted as follows: the rock has been damaged and, having an apparent density $\rho_s^0$, can be saturated by a fluid of a given true density $\hat{\rho}_f^0$ and apparent density $\rho_f^0$ without any change of deformation. Therefore, in what follows the densities $(\rho_s^0, \rho_f^0)$ are assumed to be constitutive quantities describing the state of damage of the considered rock. Therefore, we assume for $\varepsilon$ the following expansion:

$$\varepsilon(\rho_s, \rho_f) = \frac{1}{2}\frac{\partial^2 \varepsilon}{\partial \rho_s^2}\bigg|_{\rho_s^0, \rho_f^0}(\rho_s - \rho_s^0)^2 + \frac{1}{2}\frac{\partial^2 \varepsilon}{\partial \rho_f^2}\bigg|_{\rho_s^0, \rho_f^0}\left(\rho_f - \rho_f^0\right)^2$$
$$+ \frac{\partial^2 \varepsilon}{\partial \rho_f \partial \rho_s}\bigg|_{\rho_s^0, \rho_f^0}(\rho_s - \rho_s^0)\left(\rho_f - \rho_f^0\right) + O(3) \ . \tag{49}$$

A constant and a linear term are omitted to avoid pre-stresses in the natutal reference configuration. Moreover, the reference state is assumed to be saturated, so that

$$\rho_s^0 \hat{\rho}_f^0 + \rho_f^0 \hat{\rho}_s^0 = \hat{\rho}_f^0 \hat{\rho}_s^0 \ . \tag{50}$$

Note, that in a fluid filled cavern of a salt-rock formation the applied fluid pressure in the cavern generates openings of the grain boundaries between the salt crystallites in the immediate vicinity of the cavern wall. As a result, its permeability and drainage properties change. For more details see Chapter V of SMRI 1998 Technical Class [22]. We do not model here this damage process but deal only with an equilibrium situation.

As a consequence, we have for the pressures $\mathscr{P}_a$ the expansions

$$\mathscr{P}_s(\rho_s, \rho_f) = \rho_s^0 \frac{\partial^2 \varepsilon}{\partial \rho_s^2}\bigg|_{\rho_s^0, \rho_f^0}(\rho_s - \rho_s^0) + \rho_s^0 \frac{\partial^2 \varepsilon}{\partial \rho_f \partial \rho_s}\bigg|_{\rho_s^0, \rho_f^0}(\rho_f - \rho_f^0) + O(2),$$
$$\mathscr{P}_f(\rho_s, \rho_f) = \rho_f^0 \frac{\partial^2 \varepsilon}{\partial \rho_f^2}\bigg|_{\rho_s^0, \rho_f^0}(\rho_f - \rho_f^0) + \rho_f^0 \frac{\partial^2 \varepsilon}{\partial \rho_f \partial \rho_s}\bigg|_{\rho_s^0, \rho_f^0}(\rho_s - \rho_s^0) + O(2) \ , \tag{51}$$

which we will write in the compact form

$$\mathscr{P}_a = \sum_b A_{ab}(\rho_b - \rho_b^0) \ . \tag{52}$$

For compressible constituents, we assume the linear relation

$$\hat{\rho}_b = \hat{\rho}_b^0 + c_b(\rho_b - \rho_b^0) \ , \tag{53}$$

where the positive material constant $c_b$ represents the compressibility of the constituent $b$. It vanishes for a density-preserving constituent.

The constitutive part of volume exchange force

$$\sum_b \frac{\mathscr{P}_b}{v_b}(1 - \xi_b)\left(1 - v_b \frac{\mathrm{d}\hat{\rho}_b}{\mathrm{d}\rho_b}\right)^{-1} \tag{54}$$

can now be linearized: we will introduce the denotation

$$M = p + M_0 + \sum_b M_b(\rho_b - \rho_b^0) \ , \tag{55}$$

where the constants $M_0$ and $M_b$ are easily defined in terms of the constants $A_{ab}, c_b, \hat{\rho}_b^0$ and $\rho_b^0$. Their expressions are rather long and will be omitted here.

## 7
**A one-dimensional pressure-driven fluid penetration problem**

In this Section, we consider a homogeneous solid matrix saturated by a fluid occupying the halfspace $x > 0$. We will assume that all considered fields depend only on the variable $x$, that the potential $\phi$ vanishes, and the boundary of the solid matrix is located at $x = 0$. Here, the saturated solid matrix is in contact with the saturating fluid, which is kept at a fixed pressure: the properties of the interface between the saturated solid matrix and the fluid are constitutively modelled by the coefficients $\alpha$ and $l$ appearing in (46) and introduced in [20], and by the coefficient $\mathfrak{D}$ introduced in the following formula $(56)_3$. These coefficients, respectively, determine the part of the fluid pressure transmitted to solid matrix, to the saturating fluid and the contact double force exerted by the fluid on the solid matrix. The boundary conditions at $x = \infty$ are obtained by assuming that, at this boundary, the saturated solid matrix is in contact with an impervious solid.

## 7.1
### The general case of compressible constituents

The boundary conditions (45) at $x = 0$ and at $x = \infty$, implied by (46) and (47), and valid when $\alpha = 1$ and $l = 1$, are, respectively, at $x = 0$

$$-\left[\mathscr{P}_s + pv_s\left(1 - v_s\frac{d\hat{\rho}_s}{d\rho_s}\right)\right] + \lambda_s\rho_s\frac{d^2\rho_s}{dx^2} - \frac{\lambda_s}{2}\left(\frac{d\rho_s}{dx}\right)^2 = -v_s(0)p_i,$$

$$\mathscr{P}_f + pv_f\left(1 - v_f\frac{d\hat{\rho}_f}{d\rho_f}\right) = v_f(0)p_i, \qquad (56)$$

$$\lambda_s\rho_s\frac{d\rho_s}{dx} = \mathfrak{D}p_i \;,$$

and at $x = \infty$

$$-\left[\mathscr{P}_s + pv_s\left(1 - v_s\frac{d\hat{\rho}_s}{d\rho_s}\right)\right] + \lambda_s\rho_s\frac{d^2\rho_s}{dx^2} - \frac{\lambda_s}{2}\left(\frac{d\rho_s}{dx}\right)^2 = -v_s(\infty)p_i,$$

$$\left[\mathscr{P}_f + pv_f\left(1 - v_f\frac{d\hat{\rho}_f}{d\rho_f}\right)\right] = v_f(\infty)p_i, \qquad (57)$$

$$\lambda_s\rho_s\frac{d\rho_s}{dx} = 0 \;.$$

In the considered instance, adding $(44)_{1,2}$ together and integrating the emerging equation yields

$$p\left(1 - \sum_a v_a^2\frac{d\hat{\rho}_a}{d\rho_a}\right) = -\sum_a \mathscr{P}_a + \left[\lambda_s\rho_s\frac{d^2\rho_s}{dx^2} - \frac{\lambda_s}{2}\left(\frac{d\rho_s}{dx}\right)^2\right] + p_i \;, \qquad (58)$$

where for determining the integration constant the boundary conditions $(56)_1$ and $(56)_2$ were used. On the other hand, Eq. (15) for $a = f$ admits the following first integral:

$$\frac{\partial\varepsilon}{\partial\rho_f} + \frac{p}{\hat{\rho}_f}\left(1 - \frac{\rho_f}{\hat{\rho}_f}\frac{d\hat{\rho}_f}{d\rho_f}\right) = c \;, \qquad (59)$$

where $c$ is an integration constant. Eliminating $p$ from Eqs. (58) and (59) yields an ordinary differential equation, which posseses the normal form

$$\lambda_s\rho_s\frac{d^2\rho_s}{dx^2} - \frac{\lambda_s}{2}\left(\frac{d\rho_s}{dx}\right)^2 = F(\rho_s, \rho_f, c) + p_i \;, \qquad (60)$$

$$F(\rho_s, \rho_f, c) \equiv \sum_a \mathscr{P}_a + \left(1 - \sum_a v_a^2\frac{d\hat{\rho}_a}{d\rho_a}\right)\left(1 - \frac{\rho_f}{\hat{\rho}_f}\frac{d\hat{\rho}_f}{d\rho_f}\right)^{-1}\left(c - \frac{\partial\varepsilon}{\partial\rho_f}\right)\hat{\rho}_f \;. \qquad (61)$$

Differentiating (60) yields

$$\frac{d}{dx}\left[\lambda_s\rho_s\frac{d^2\rho_s}{dx^2} - \frac{\lambda_s}{2}\left(\frac{d\rho_s}{dx}\right)^2\right] = \lambda_s\rho_s\frac{d}{dx}\left(\frac{d^2\rho_s}{dx^2}\right) = \frac{d}{dx}F(\rho_s, \rho_f, c) \;, \qquad (62)$$

which implies

$$\lambda_s \frac{d^2 \rho_s}{dx^2} = G(\rho_s, c) + g_1(c, p_i, \mathfrak{D}) \ , \tag{63}$$

where

$$G(\rho_s, c) := \int \frac{1}{\rho_s} \frac{d}{d\rho_s} F[\rho_s, \bar{\rho}_f(\rho_s), c] df_s \tag{64}$$

and the function $\bar{\rho}_f(\cdot)$ is implicitly defined by the saturation constraint

$$\rho_s \hat{\rho}_f(\rho_f) + \rho_f \hat{\rho}_s(\rho_s) = \hat{\rho}_f(\rho_f) \hat{\rho}_s(\rho_s) \ . \tag{65}$$

The integration constants $c$ and $g_1(c, p_i, \mathfrak{D})$, together with the further two integration constants of (63), are determined by using the boundedness of the solution (57) as $x \to \infty$, and (56)$_{2,3}$ for $x = 0$.

## 7.2
## Incompressible constituents

To show some of the qualitative features implied by (63)–(65) we assume that

$$\frac{d\hat{\rho}_a}{d\rho_a} = 0, \quad \rho_s \hat{\rho}_f^0 + \rho_f \hat{\rho}_s^0 = \hat{\rho}_f^0 \hat{\rho}_s^0 \ . \tag{66}$$

The boundary conditions at $x = 0$ are

$$-\left[ A_{ss}(\rho_s - \rho_s^0) + A_{sf}(\rho_f - \rho_f^0) + p \frac{\rho_s}{\hat{\rho}_s^0} \right] + \lambda_s \rho_s \frac{d^2 \rho_s}{dx^2} - \frac{\lambda_s}{2} \left( \frac{d\rho_s}{dx} \right)^2 = -\frac{\rho_s(0)}{\hat{\rho}_s^0} p_i,$$

$$A_{ff}(\rho_f - \rho_f^0) + A_{fs}(\rho_s - \rho_s^0) + p\left(1 - \frac{\rho_s}{\hat{\rho}_s^0}\right) = \left(1 - \frac{\rho_s(0)}{\hat{\rho}_s^0}\right) p_i, \tag{67}$$

$$\lambda_s \rho_s \frac{d\rho_s}{dx} = \mathfrak{D} p_i \ ,$$

and Eq. (58) reduces to

$$p = -\sum_a \mathscr{P}_a + \lambda_s \rho_s \frac{d^2 \rho_s}{dx^2} - \frac{\lambda_s}{2} \left( \frac{d\rho_s}{dx} \right)^2 + p_i \ . \tag{68}$$

Replacing the expression for $p$ in (67)$_2$, we get at $x = 0$

$$A_{ff}(\rho_f - \rho_f^0) + A_{fs}(\rho_s - \rho_s^0) + \left\{ -\sum_a \mathscr{P}_a + \left[ \lambda_s \rho_s \frac{d^2 \rho_s}{dx^2} - \frac{\lambda_s}{2} \left( \frac{d\rho_s}{dx} \right)^2 \right] \right\} \left(1 - \frac{\rho_s}{\hat{\rho}_s^0}\right) = 0 \ . \tag{69}$$

On the other hand, Eq. (59) takes the form

$$\frac{A_{ff}}{\rho_f^0}(\rho_f - \rho_f^0) + \frac{A_{fs}}{\rho_f^0}(\rho_s - \rho_s^0) + \frac{p}{\hat{\rho}_f^0} = c \ . \tag{70}$$

By eliminating $p$ from (68) and (70), we obtain the ordinary differential equation for $\rho_s$

$$\left[ -\frac{A_{ff}}{\rho_f^0}(\rho_f - \rho_f^0) - \frac{A_{fs}}{\rho_f^0}(\rho_s - \rho_s^0) + c \right] \hat{\rho}_f^0$$

$$= -A_{ff}(\rho_f - \rho_f^0) - A_{ss}(\rho_s - \rho_s^0) + \left[ \lambda_s \rho_s \frac{d^2 \rho_s}{dx^2} - \frac{\lambda_s}{2} \left( \frac{d\rho_s}{dx} \right)^2 \right] + p_i \ , \tag{71}$$

which has the normal form

$$\lambda_s \rho_s \frac{d^2\rho_s}{dx^2} - \frac{\lambda_s}{2}\left(\frac{d\rho_s}{dx}\right)^2 = \left[A_{ff}\left(1 - \frac{\hat{\rho}_f^0}{\rho_f^0}\right) + A_{sf}\right](\rho_f - \rho_f^0)$$
$$+ \left[A_{ss} - A_{fs}\frac{\hat{\rho}_f^0}{\rho_f^0}\right](\rho_s - \rho_s^0) - p_i + c\hat{\rho}_f^0 =: F(\rho_s, \rho_f, c) \ . \tag{72}$$

Because of (62)$_2$, it implies

$$\lambda_s \frac{d^2\rho_s}{dx^2} = L \ln \frac{\rho_s}{\rho_s^0} + g \simeq L\left(\frac{\rho_s - \rho_s^0}{\rho_s^0}\right) + g = L\frac{\rho_s}{\rho_s^0} - (L - g) \ , \tag{73}$$

where

$$L \equiv A_{ss} - A_{fs}\frac{\hat{\rho}_f^0}{\rho_f^0} - \left[A_{ff}\left(1 - \frac{\hat{\rho}_f^0}{\rho_f^0}\right) + A_{sf}\right]\frac{\hat{\rho}_f^0}{\rho_s^0} \ . \tag{74}$$

It possesses the dimension of energy per unit mass, and may, according to its definition, be positive or negative, depending on the relative magnitude of the coefficients $A_{ab}$. This may give way to a branching solution. However, in anticipation that $A_{ss}$ is much larger than $A_{ff}$, $A_{sf}$ or $A_{fs}$, $L$ is likely positive, and the solution of the ordinary differential equation (73) is decaying in $x$. In this spirit, therefore,

$$\rho_s(x) = \rho_s^0\left(\frac{L-g}{L}\right) + C_1 e^{\sqrt{\frac{L}{\rho_s^0 \lambda_s}}x} + C_2 e^{-\sqrt{\frac{L}{\rho_s^0 \lambda_s}}x} \ . \tag{75}$$

Now we impose that

1. At $x \to \infty$, the boundary conditions (57) hold. In particular, (57)$_3$ implies that

$$C_1 = 0 \Rightarrow \left.\frac{d^2\rho_s}{dx^2}\right|_{x=\infty} = 0 \quad \text{and} \quad \left.\frac{d\rho_s}{dx}\right|_{x=\infty} = 0 \ , \tag{76}$$

and, therefore, with simple algebra, (57)$_{1,2}$ become

$$\mathscr{P}_a|_{x=\infty} = 0 \Rightarrow \rho_a(\infty) = \rho_a^0 \quad (a = s, f) \Rightarrow g = 0 \ . \tag{77}$$

In this deduction one also must use $\hat{p}(\infty) = p_i$, which is the result of global equilibrium.
2. At $x = 0$,

$$\frac{d\rho_s(0)}{dx} = \frac{\mathfrak{D}p_i}{\lambda_s \rho_s(0)} \ , \tag{78}$$

which implies

$$C_2(\rho_s^0 + C_2) = -\mathfrak{D}p_i\sqrt{\frac{\rho_s^0}{\lambda_s L}} \ . \tag{79}$$

Because of the linearization procedure, which we have used, the only meaningful solution for $C_2$ is given by

$$\frac{C_2}{\rho_s^0} = -\frac{1}{2} + \frac{1}{2}\sqrt{1 - 4\left(\frac{\mathfrak{D}p_i}{(\rho_s^0)^2}\sqrt{\frac{\rho_s^0}{\lambda_s L}}\right)} \simeq -\frac{\mathfrak{D}p_i}{(\rho_s^0)^2}\sqrt{\frac{\rho_s^0}{\lambda_s L}} \ . \tag{80}$$

In conclusion we have

$$\rho_s(x) = \rho_s^0 - \frac{\mathfrak{D}p_i}{\rho_s^0}\sqrt{\frac{\rho_s^0}{\lambda_s L}}e^{-\sqrt{\frac{L}{\rho_s^0 \lambda_s}}x} \quad . \tag{81}$$

## 8
## Static permeability of porous solid matrices and conclusions

Formula (81) shows that the second-gradient theory, which has been introduced in this paper, features important differences when compared with the classical theory of Fillunger and Terzaghi.

Indeed, when $\lambda_s = 0$ the apparent solid density is constant and independent of the pressure $p_i$. On the other hand, when the second-gradient strain has an influence on the stress, and, hence, $\lambda_s \neq 0$, then (81) shows that the length

$$x_0 := \sqrt{\frac{\rho_s^0 \lambda_s}{L}} \quad , \tag{82}$$

plays the role of an attenuation length for the apparent solid mass density.

We will, henceforth, call the quantity $x_0$ static permeability of the solid matrix as it is this characteristic length of the exponential decay (81) that measures the amount of saturating fluid which has penetrated into the solid porous matrix under equilibrium conditions. It is evident that the introduced static permeability is quite different from the dynamic Darcy permeability, which is defined in terms of relative velocities driven by pressure jumps or gradients.

The density variation

$$\Delta\rho_s := \frac{\mathfrak{D}p_i}{\rho_s^0}\sqrt{\frac{\rho_s^0}{\lambda_s L}} \quad , \tag{83}$$

which depends on the contact double-force coefficient $\mathfrak{D}$, measures the intensity of the $p_i$-induced drop of the apparent solid mass density, and can be called static permeability of the solid matrix boundary.

To conclude, we remark that:

1. The expression (58), which is valid under equilibrium conditions, shows that in the generalized model the saturation pressure can no longer be interpreted immediately as a porewater pressure; indeed, under equilibrium conditions, the fluid pressure in a connected region is constant but $p$ is not.
2. Again (58) shows that the compressibility of the saturating fluid and solid matrix greatly influence the values of the saturation pressure.
3. The expressions for static permeabilities, which are found in this Section, are valid only for the case of incompressible solid and fluid constituents.
4. Although found for a one-dimensional case and, therefore, apparently far from engineering applications, formula (81) gives a first interesting applicable answer to the problems raised recently in [22–24].

**References**
  1. **de Boer, R.; Ehlers, W.:** The development of the concept of effective stresses. Acta. Mech. 83 (1990) 77–92
  2. **Germain, P.:** La méthode des puissances virtuelles en mécanique des milieux continus. Premiè partie: Théorie du second gradient. Journal de Mécanique 12 (1972) 235–274
  3. **Germain, P.:** The method of virtual power in continuum mechanics Part 2: Microstructure. S.I.A.M. J. Appl. Math. 25 (1973) 556–575
  4. **Svendsen, B.:** A thermodynamic model for volume-fraction-based mixtures. Z. Angew. Math. Mech. 77S2 (1997) S413–S416
  5. **Svendsen, B.; Hutter, K.:** On the thermodynamics of a mixture of isotropic materials with constraints. Int. J. Eng. Sci. 33 (1995) 2021–2054
  6. **Krishnaswamy, S.; Batra, R. C.:** A thermomechanical theory of solid-fluid mixtures. Math. Mech. Solids 2 (1997) 143–151

7. **Seppecher, P.:** Etude des conditions aux limites en théorie du second-gradient: cas de la capillarité. C. R. Acad. Sci. S. II 309 (1989) 497–502
8. **dell'Isola, F.; Seppecher, P.:** The relationship between edge contact forces, double forces and interstitial working allowed by the principle of virtual power. C. R. Acad. Sci. S. IIb 321 (1995) 303–308
9. **dell'Isola, F.; Seppecher, P.:** Edge contact forces and quasi-balanced power. Meccanica 32 (1997) 33–52
10. **Berest, P.; Brouard, B.:** A tentative classification of salts according to their creep properties SMRI Proceedings Spring Meeting, New Orleans, USA 1998
11. **Biot, M.:** General theory of three-dimensional consolidation. J. Appl. Physics 12 (1941) 648–657
12. **Rice, J. R.; Cleary, M. P.:** Some basic stress diffusion solutions for fluid-saturated elastic porous media with compressible constituents. Reviews of Geophysics and Space Physics 14 (1976) 227–241
13. **Coussy, O.:** Mecanique des milieux poreux, Technip. 1991
14. **Ehlers, W.:** Constitutive equations for granular materials in geomechanical context. In: K. Hutter (ed.) Continuum Mechanics in Environmental Sciences and Geophysics, 313–402 Wien, Springer-Verlag 1993
15. **Drew, D. A.; Segel, L. A.:** Averaged equations for two phase flows. Stud. Appl. Math. L (1971) 205–231
16. **Drew, D. A.; Lahey, R. T.:** Application of general constitutive principles to the derivation of multidimensional two-phase flow equations Int. J. Multiphase Flow 5 (1979) 243–262
17. **Pfeifle, T. W.; DeVries, K. L.; Nieland, J. D.:** Damage-Induced Permeability Enhancement of Natural Rock Salt with Implications for Cavern Storage. SMRI Spring Meeting, New Orleans, 1998
18. **Seppecher, P.:** Etude d'une modelisation des zones capillaires fluides: interfaces et lignes de contact. These Univ. Paris VI 1987
19. **dell'Isola, F.; Hutter, K.:** What are the dominant thermomechanical processes in the basal sediment layer of large ice sheets? Proc. R. Soc. London A 454 (1998) 1169–1195
20. **dell'Isola, F.; Hutter, K.:** A qualitative Analysis of the dynamics of sheared and pressurized layer of saturated soil. Proc. R. Soc. London A 454 (1998) 3105–3120
21. **dell'Isola, F.; Hutter, K.:** Destabilization of a sheared pressurized layer of soil by drainage of water to appear in Proc. R. Soc. London A (1999)
22. **SMRI Solution Mining Research Institute:** Fall Meeting. Technical Class Guidelines for Safety Assessment of Salt Caverns, Rome 1998
23. **Berest, P.; Brouard, B.; Bergues, J.; Frelat, J.; Durup, J. G.:** Salt caverns and the compressibility factors, SMRI Proceedings. Fall Meeting El Paso 1997
24. **SMRI Solution Mining Research Institute:** Fall Meeting, Rome Proceedings 1998
25. **Fillunger, P.:** Erdbaumechanik Wien: Selbstverlag des Verfassers 1936